\documentclass{JAC2003}

\usepackage{graphicx}
\usepackage{booktabs}

\begin{document}
\title{Instrumentation2: Other instruments, ghost/satellite bunch monitoring, halo, emittance, new developments\thanks{This contribution is presented by the author on behalf of the BE-BI group}}

\author{E. Bravin, CERN, Geneva, Switzerland}

\maketitle

\begin{abstract}
In order to estimate in absolute terms the luminosity of LHC certain beam parameters have to be measured very accurately. In particular the total beam current and the relative distribution of the charges around the ring, the transverse size of the beams at the interaction points and the relative position of the beams at the interaction point.
The experiments can themselves measure several of these parameters very accurately thanks to the versatility of their detectors, other parameters need however to be measured using the monitors installed on the machine.
The beam instrumentation is usually built for the purpose of aiding the operation team in setting up and optimizing the beams, often this only requires precise relative measurements and therefore the absolute scale is usually not very precisely calibrated. The luminosity calibration requires several machine-side instruments to be pushed beyond their initial scope.
\end{abstract}

\section{Colliding and non colliding charges}
In general in colliders the particles circulating in opposite directions are kept separated and only allowed to encounter each other at the designated interaction points. This is even more true for the LHC where the particles travel in different vacuum tubes for most of the accelerator length. Particles colliding outside of the experiments would provide no useful information and would only contribute to the background and reduce the lifetime of the beams. In order to estimate the luminosity it is therefore important to quantify the number of particles that can potentially collide in a given interaction point more than just the total current stored in the machine.
The distribution of particles around the ring can be rather complicated. In theory there should be only a well known number of equal bunches spaced by well known amounts of time and in this situation it would be easy to calculate the colliding charges from the total current. In reality the bunches have all different currents and there can be charges also outside of these bunches. In the LHC the radio frequency (RF) system has a frequency of 400.8~MHz and only every 10th bucket at most is filled. This means that there are plenty of \emph{wrong} RF buckets that can store particles in a stable way. It can happen that capture problems (also upstream in the injectors) create unwanted small intensity bunches near by the main ones. These, named satellite bunches, have typically intensities of up to 1\% of the main bunch and are only a few RF buckets away from the main bunch (usually a multiple of the RF period of one of the preceding accelerators).  Other effects can lead to particles escaping from the main buckets and becoming un-captured, these particles are no longer synchronous and will just diffuse around the ring where they can remain for very long time. In case some RF gymnastic is performed (like inserting dips in the accelerating voltage in order to improve injection efficiency) it can happen that some un-captured beam is recaptured forming a very large number of very faint bunches. These are called ghost bunches and have typically currents below the per-mill of the main bunches. In the LHC ghost bunches have been observed, in particular during the heavy ions run due to the special RF tricks used at injection when injecting ions.  It is worth mentioning that un-captured particles will be lost if the energy of the machine is changed (e.g. during the ramp) due to the fact that they can not be properly accelerated by not being synchronous with the RF.

\section{Measuring the colliding charge}
Usually fast current transformers should be sufficient to measure the relative current variations from bunch to bunch. The dynamic range and speed of these detectors are however not sufficient to detect the satellites and the ghost bunches. Moreover in the LHC the fast current transformers integrate the beam current over 25\,ns (10 RF buckets) bins and it is not possible to know if and which satellites are included in the integration. Detectors with better time resolution and higher dynamic range are required.
Candidates are:
\begin{Itemize}
    \item  Wall current monitor
    \item  Strip line pick-up
    \item  Fast light detector sampling the synchrotron light vs. time
    \item  Precise time stamping and counting of synchrotron light
\end{Itemize}

\subsection{Wall current monitor}
\begin{figure}[htb]
   \centering
   \includegraphics*[width=1.0\linewidth]{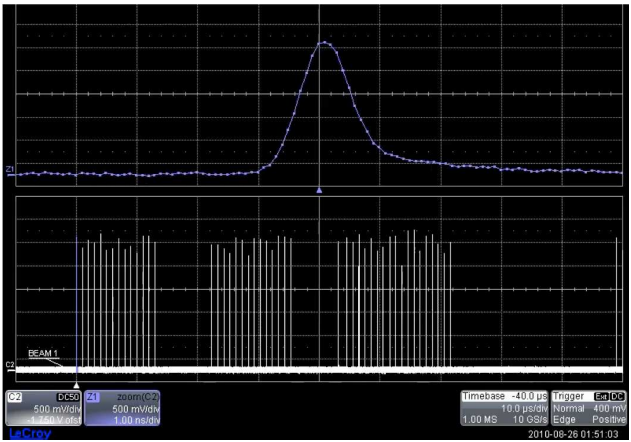}
   \caption{Signal from a wall current monitor. The top graph shows a zoom into a single bunch while the bottom graph shows the entire ring.}
   \label{WCM}
\end{figure}

The wall current monitors can probably be used to estimate the satellites. This requires however averaging over many turns and correcting for quirks in the frequency response of the detector and the cables. It is in particular important to verify that reflection/noise or other effects are not limiting the potential of the averaging. For the moment the amount of charge in satellites is calculated by studying the frequency spectra of the acquired signals, as satellites are out of the nominal bunching pattern it is possible to compare the expected spectra with the measured one ad estimate the amount of charge producing the distortion. One complication to this process arises from the fact that the bunches are not necessarily Gaussian and their shape is not precisely known. It is however difficult to get sufficient accuracy in order to take care of the ghosts. At the moment a continuous analysis of the spectrum of the wall current monitor is performed by the front-end software and provides an estimate of the amount of charge outside of the correct buckets which is stored in the database. Figure~\ref{WCM} shows the signal from a wall current monitor acquired with a 10~GSample scope. A long tail after the bunch can be observed, this arises from the frequency response of the detector and is corrected for in the analysis.

\subsection{Strip line pick-ups}
\begin{figure}[htb]
   \centering
   \includegraphics*[width=1.0\linewidth]{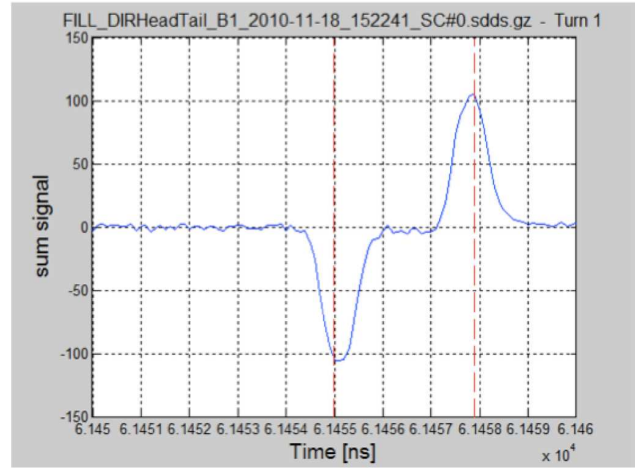}
   \caption{Signal from a strip line pick-up.}
   \label{strip_line}
\end{figure}

The strip line pick-ups provide signals comparable to the ones of the wall current monitor with the drawback of a perfectly reflected pulse shortly after the main pulse with a delay that depends on the strip length, 30\,cm for the devices installed in the LHC, intrinsic to the principle of the device (see Fig.~\ref{strip_line}). This reflection complicates the treatment of the signal resulting in the impossibility of using this instrument for the identification of ghosts and satellites.

\subsection{Synchrotron light detection}
\begin{figure*}[tb]
\centering
\begin{tabular}{c @{\hspace{1cm}} c @{\hspace{1cm}} c @{\hspace{1cm}} c}
\includegraphics[width=0.2\linewidth]{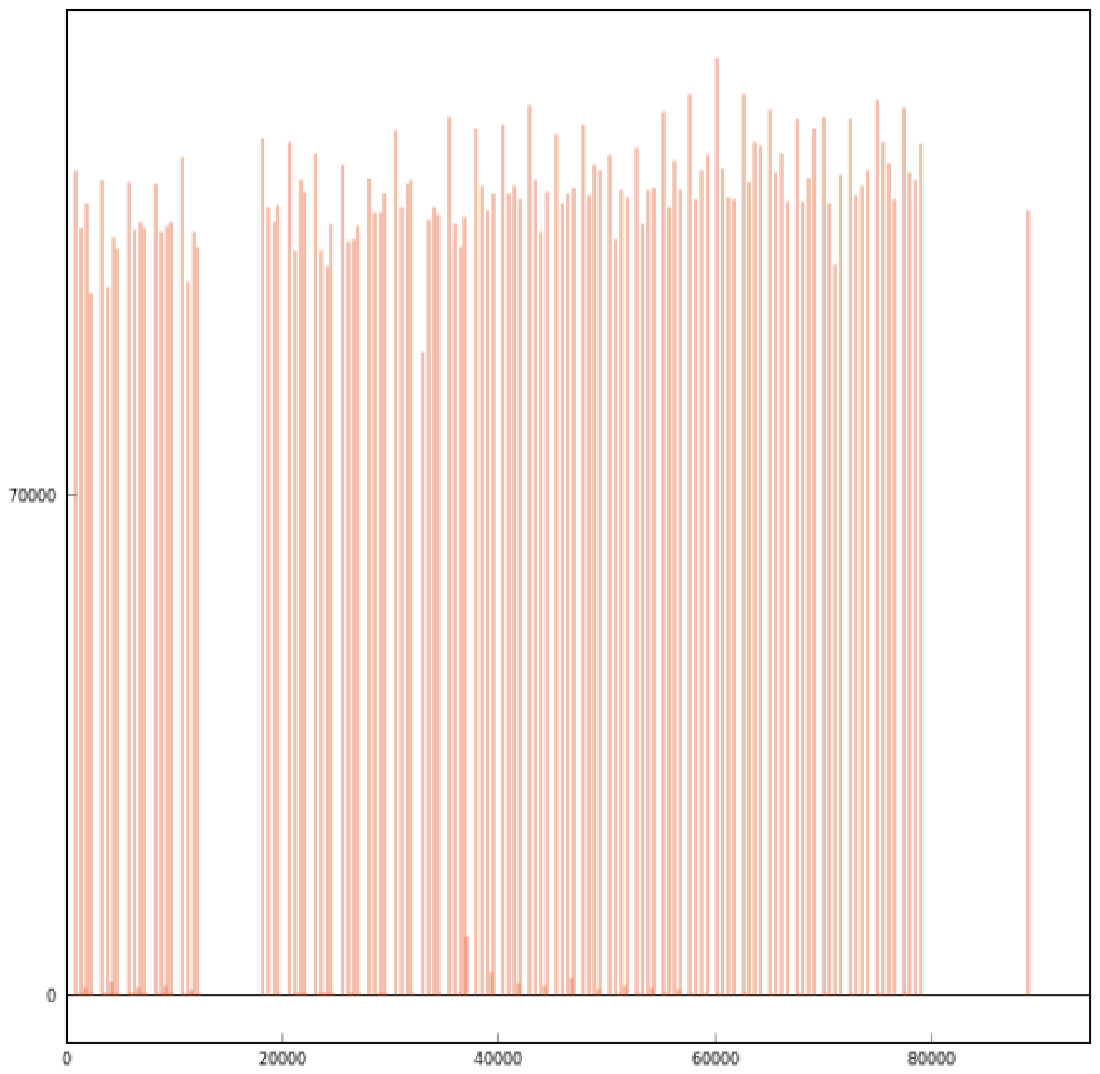} &
\includegraphics[width=0.2\linewidth]{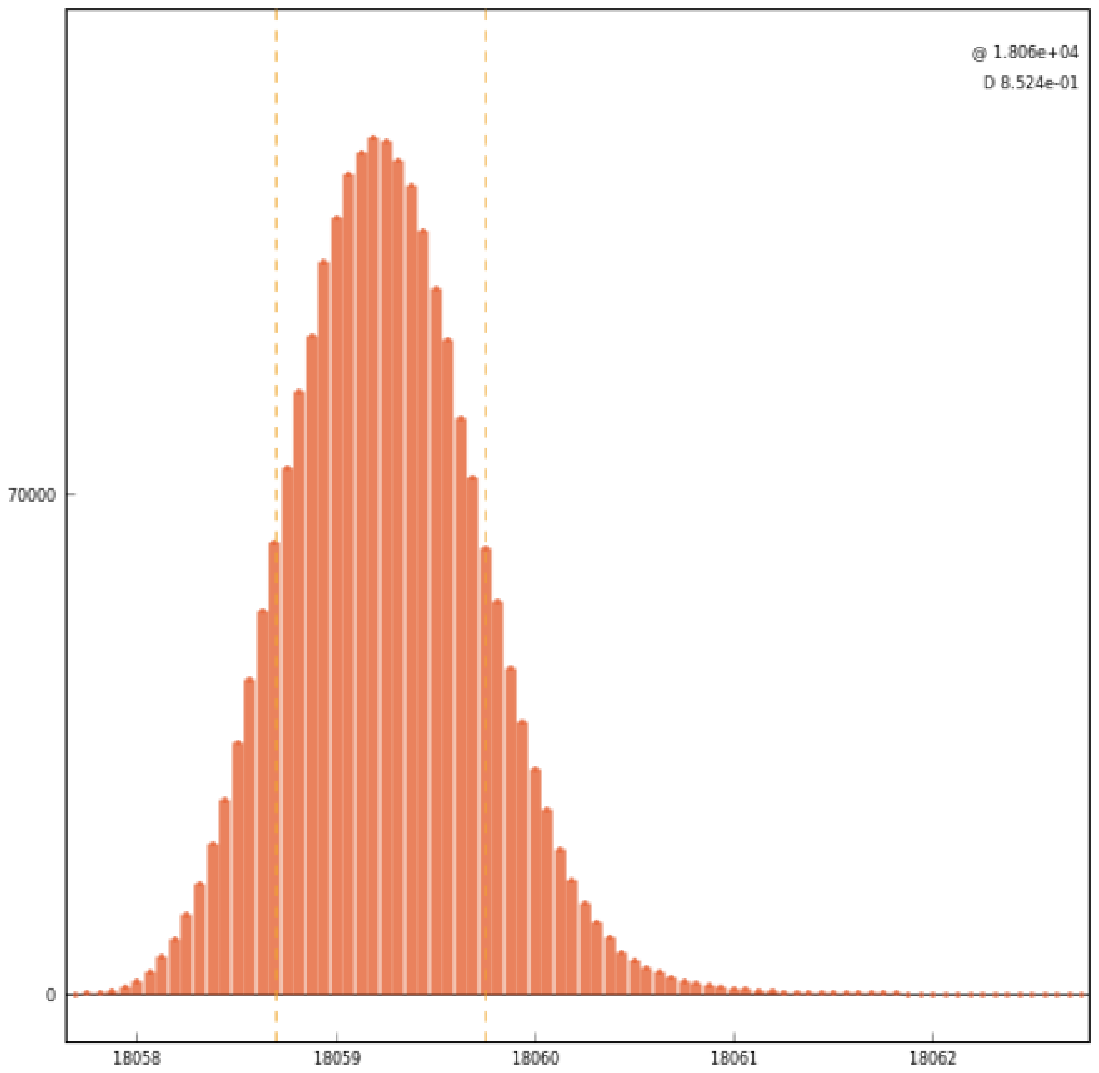} &
\includegraphics[width=0.2\linewidth]{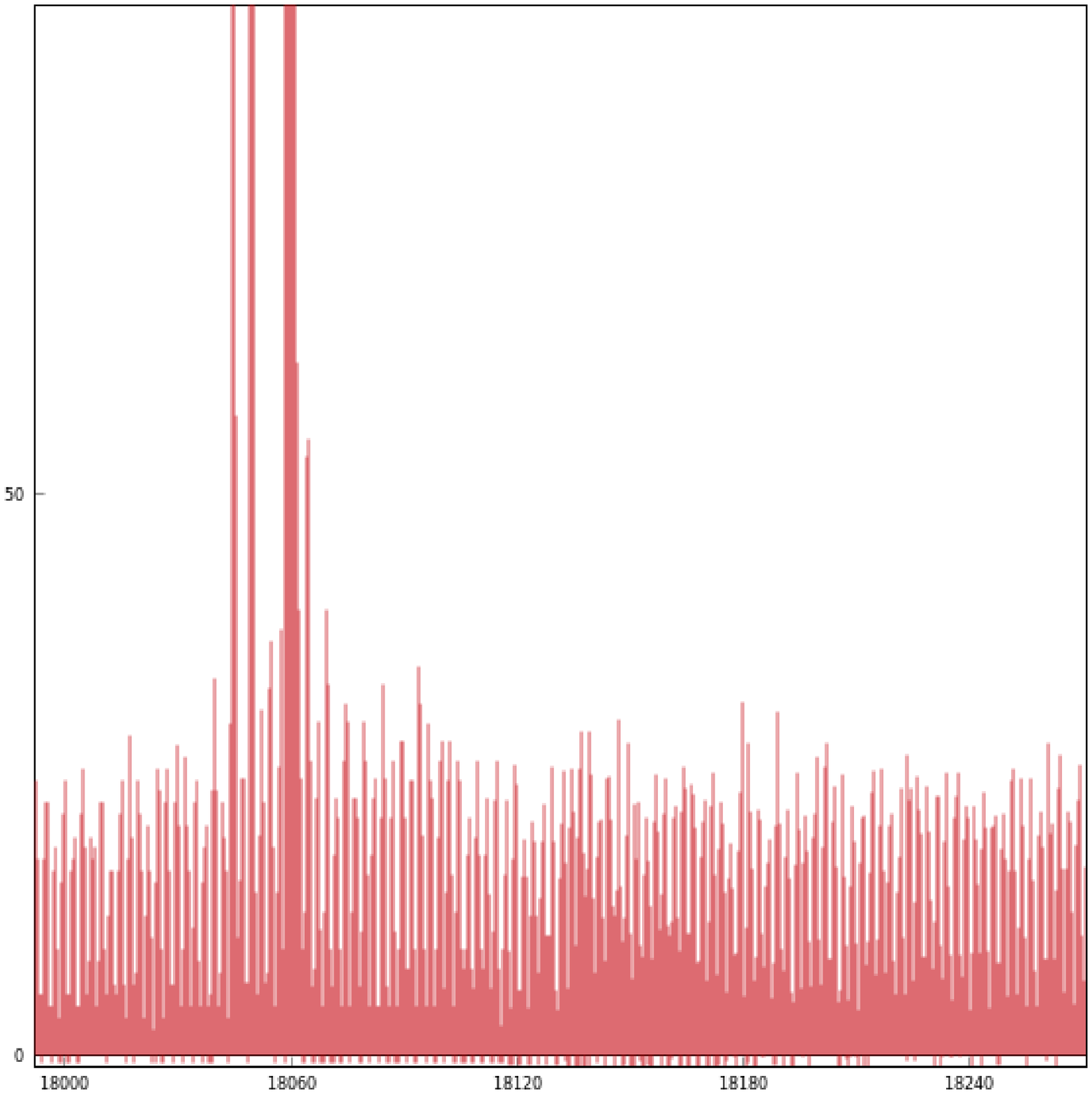} &
\includegraphics[width=0.2\linewidth]{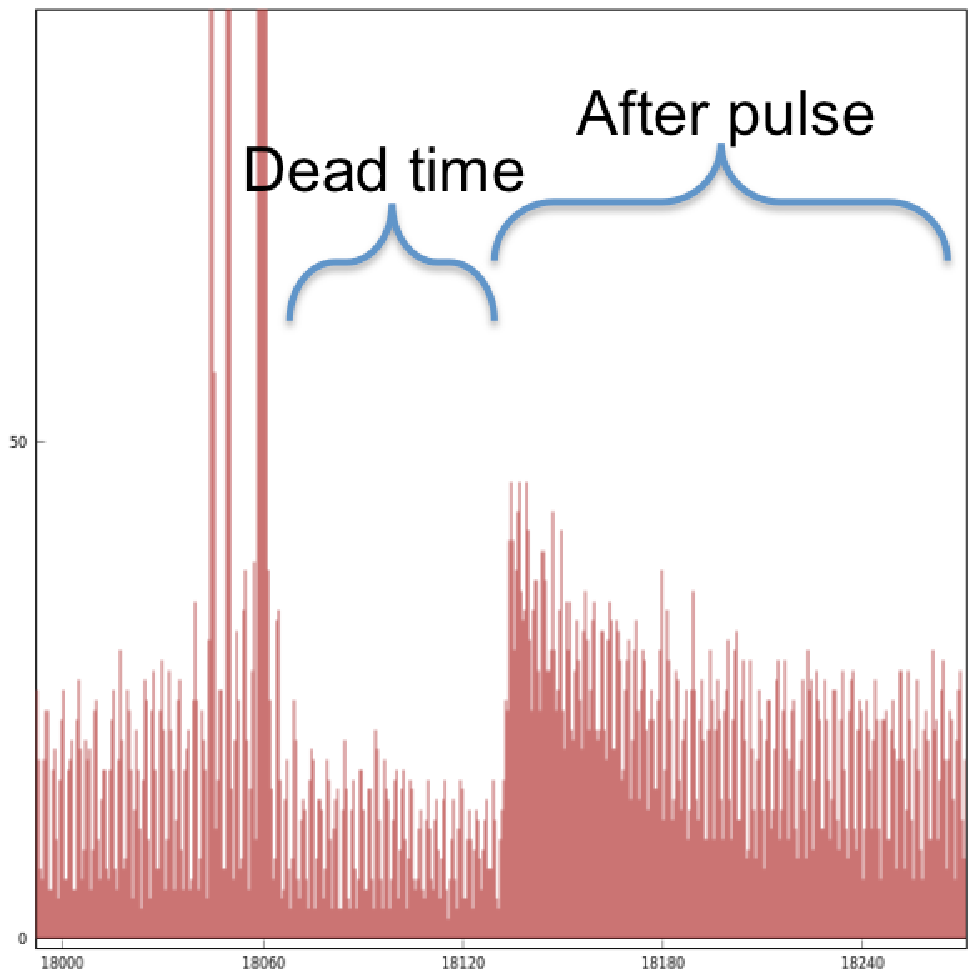} \\
a) & b) & c) & d) \\
\end{tabular}
\caption{Longitudinal profiles measured with the LDM. Distribution of particles around the whole ring (a), zoom on a single bunch (b), zoom at the base of a bunch showing the main bunch (thick line), satellites (thinner lines next to the main bunch) and ghost bunches ("noise" lines around the baseline) (c). Same zoom as (c), but before correction (d).}
\label{LDM}
\end{figure*}

There are two possibilities for using the synchrotron light for longitudinal measurements. One consists in simply using a fast optical detector connected to a fast sampler and record the intensity of synchrotron light as function of time. The principle is simple and photo-diodes in the order of 50\,GHz are commercially available, there are however a few difficulties associated with this technique. As for the WCM the transport of the high frequency signals is not simple and the cables response will modify the pulses requiring frequency domain corrections. Another problem is introduced by the need of fast digitizers implying a reduced dynamic range (typically 8 bits only), noise etc. On the other end the response of the detector itself should be much more linear than the one of the WCM and can in principle extend down to DC. It is surely worth trying this possibility however it will be very difficult to be able to measure the ghost bunches in this way.
The other alternative is to count single SR photons with precise time stamping of the arrival time. Detectors suitable for the task exists (avalanche photo diodes, APD) and time to digital converters with resolutions of a few tens of ps also exists. The only draw back of this technique is that the counting rate is limited and the light has to be attenuated such that the probability of detecting a photon during a bunch passage should be less than ~60\%.
Such a detector has been operated during the last part of the 2010 run (mainly during the ions period) and has given very promising results, it is known as the longitudinal density monitor or LDM (see Fig.~\ref{LDM}.)

\subsection{Longitudinal density monitor LDM}
The LDM is based on avalanche photodiodes from either id-Quantique or Micro Photon Devices connected to a fast TDC from Agilent (former Acquiris). The detector can resolve single photons with a time resolution of the order of 50\,ps, the TDC has a resolution of 50\,ps as well. At the moment the temporal resolution of the system is limited to about 75\,ps (300\,ps pk-pk) due to the reference timing used (turn clock from the BST receiver, BOBR), in the future this limitation will be removed by using a dedicated RF timing signal \cite{ldm}.
The avalanche photo diodes present a short dead-time used to quench the avalanche (tens of ns) and there is also a small probability that  at the end of this dead-time trapped electrons or holes will trigger a new avalanche (the probability of this type of events is of the order of 3\%.) These effects, together with the dark count rate, although small, must be corrected for, a rather simple statistical algorithms is sufficient. The probability of SL photon triggering an avalanche per bunch-crossing must be maintained below a certain level (60-70\%) otherwise the error on these corrections becomes too large. This has an impact on the maximum counting rate and thus on the integration time required for acquiring a profile with sufficient resolution.
In fact the integration time required depends on what is being observed; if the aim is just to measure the so called core parameters of a bunch (mainly the bunch length) a few seconds are sufficient, on the other hand if the population of ghosts and satellites has to be measured an integration of several minutes may be required.
The dynamic range observed in 2010 was of the order of $10^6$ with an integration time of 500~s.
The LDMs consist of an extension to the already complex synchrotron light telescope, this means that there may be interferences between the optimization of the LDM and the other detectors present on the optical table (fast and slow cameras and abort gap monitor.) In 2011 the LDM should become operational for both beams. 
\begin{figure}[htb]
   \centering
   \includegraphics*[width=1.0\linewidth]{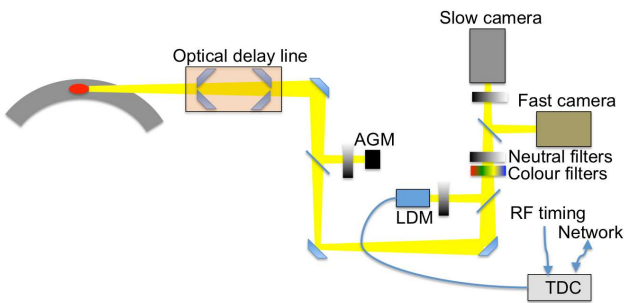}
   \caption{Schematics of the BSRT optical system.}
   \label{bsrt_layout}
\end{figure}

\section{Bunch length}
At the moment bunch lengths is LHC are typically of the order of 0.8\,ns FWHM, the nominal value is 250\,ps one sigma. In order to measure this parameter a detector with high bandwidth is required (several GHz) however even a limited dynamic range would be sufficient. The list of candidates for this measurement is similar to the one presented before for the measurement of the satellites/ghost bunches
\begin{Itemize}
    \item  Wall current monitor
    \item  Strip line pick-up
    \item  Fast light detector sampling the synchrotron light vs. time
    \item  LDM
\end{Itemize}

\subsection{Wall current monitor}
This device measures the image current flowing on the beam pipe. The WCMs installed on LHC have an upper cut-off frequency of about 3\,GHz and the signals are sampled using a scope with 10\,GSample/s. These characteristics are sufficient for the measurement of the bunch length, however the non flat transfer function of the detectors introduce tails at the end of the bunch. By analyzing the signals in frequency domain these artifacts can be removed, Fig.~\ref{WCM} shows the signal directly on the scope display before processing. 

\subsection{Strip line pick-up}
The main function of this device is to measure the position of the beam with high temporal accuracy, in particular it allows to study the head-tail oscillations of the beam which provide hindsights on the stability of the beams and also a way to measure the chromaticity (variation of the betatron tune vs. the error in momentum of the particle).
The device is composed of 4 electrodes, 30\,cm long, mounted at 90${}^\circ$. The amplitude of the signal on each electrode depends on the instantaneous beam current as well of the distance between the bunch and the electrode. By summing the signals on opposite electrodes one obtains a signal only proportional to the beam current while subtracting the signals from opposite electrodes and dividing by the sum one obtains a signal proportional to the position only. The bandwidth is similar to the one of the WCM, mainly limited by the characteristics of the feed-through and resonances in the electrodes. The acquisition is in fact performed with the same type of scope used for the WCM. The advantage of the strip line is that the transfer function is almost flat. Another characteristic of the strip line detectors has been already mentioned and consists of a second pulse, inverted in polarity, after the first one, the distance between the two being determined by the length of the electrodes (to be precise twice the length of the line divided by the speed of light) see Fig.~\ref{strip_line}.
Of course both the WCM and the strip line can provide single passage measurements as well as averaged measurements.

\subsection{LDM}
As seen before the LDM allows the sampling of the whole LHC ring with high time accuracy, with the present system 50\,ps resolution is possible. The 50\,ps temporal resolution is enough to measure the bunch length, provided the beam is stable over the integration time needed to acquire a profile, typically a few seconds.
It has already been mentioned that the intensity of the synchrotron light could be acquired directly by a photon detector instead of performing single photon counting. This technique has however not yet been used in LHC as it would carry all the problematics of the strip-line pick-ups for example (cables transfer functions, fast sampling) without adding substantial advantages.

\section{Transverse emittance}
Another important factor in the determination of the luminosity from the machine parameters is the transverse emittance. Several instruments have been installed in the LHC for this purpose. In particular the instrument used to measure accurately the beam size and thus the beam emittance is the wire scanner. This instruments however only produces measurements on demand and can not be used when the total beam intensity is above 2~$10^{13}$\,protons. 
In order to cope with the limitations of the wire scanner two different monitors capable of continuous monitoring have been installed, the synchrotron light telescope (BSRT) and the rest gas ionization monitor (BGI.)
All these devices only measure the transverse beam sizes, in order to calculate the emittance the knowledge of the optics of the machine at the location of the devices is needed, in particular the betatron function. Thanks to the accurate modeling and precise measurements the beta functions are known with an error between 5 and 10\% all around the machine.

\subsection{Wire scanner}
This is the reference device for emittance measurement, since the systematic errors of this technique can be controlled well. The principle is rather simple and consists of scanning a 30\,$\mu$m diameter carbon wire across the beam at about 1\,m/s. The interaction of the particles in the beam with the nuclei in the wire produce high energetic secondary particles that are detected by a scintillator-photomultiplier assembly some 10\,m downstream of the scanner. The beam profile is inferred by the amplitude of the PMT signal as function of the wire position.
Because the wire scanner needs to intercept the beam in order to make a measurement the range of beam intensity were it can be used is limited. There are two situations that need to be avoided: overheating the wire up to the point were it breaks or inducing secondary particles and beam losses of intensity sufficient to induce a quench in the neighboring superconducting magnets. At injection energy the first effect dominates while at top energy it is the second. The intensities of these two limits are rather close and for this reason only one value (the smaller) is used imposing an upper limit of 2~$10^{13}$\,particles per beam (about 200 nominal bunches) \cite{ws_limits}.
The accuracy of the wire scanner in the LHC has not yet been studied, however a detailed study on similar devices has been carried out in the SPS a few years ago leading to an error of the order of 1\% in the beam emittance for beams of $\sigma=1$\,mm transverse size \cite{ws_accuracy}. At the end of 2010 the bunch-by-bunch acquisition mode was also commissioning, the wire scanners can thus be used now either to measure the average over all bunches or the profile of individual bunches.

\subsection{BSRT}
The two synchrotron light telescopes are installed at point 4 and take advantage of the D3 dipoles used to separate the beam around the RF cavities. Since at injection energy the spectra from these dipoles is in the far infrared two undulators have been developed and installed at the entrance of the D3. These special magnets provide sufficient radiation in the visible range up to about 1\,TeV where the radiation from the dipole magnet takes over, Fig.~\ref{bsrt_layout} shows a simplified sketch of the BSRT setup. The imaging requires a complex mirror-based optical telescope and since it can not be accessed when there is beam present many components have to be adjusted remotely.
The image acquisition is performed by an intensified camera which image intensifier can be gated to a single bunch allowing single bunch single turn acquisition. By scanning the gate delay all the bunches can be scanned in turn, this process is however long since the acquisition system is limited to one image per second. This type of bunch scan was performed regularly at the end of the 2010 run. Another camera, the fast camera, allows the single bunch single turn measurement, but in this case images can be acquired at over 11\,kHz allowing the acquisition of the full ring in a fraction of second. The fast camera was not installed in 2010, but will be made operational during 2011.
Due to the complexity of the optical system and the many constrains the optical resolution of the telescope is intrinsically limited to a few hundred microns \cite{bsrt_alan}, this limit has not yet been achieved yet and the reasons are not entirely understood \cite{bsrt_first}. The point spread function of the BSRTs have been calculated by comparing the sizes measured by the BSRT and the wire scanners, this PSFs are then de-convoluted from the measured values
\begin{equation}\label{eq:PSF}
    \sigma_{beam}=\sqrt{\sigma_{meas}^2-PSF^2}
\end{equation}
Presently the PSF values are different for the two beams and for the two planes, but are all around 500\,$\mu$m.

\begin{figure}[htb]
\centering
\includegraphics[width=1.0\linewidth]{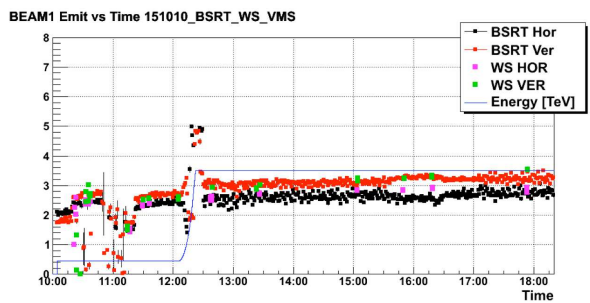} 
\caption{Evolution of the B1 beam emittance during a fill as measured by the BSRT and the wire scanner}
\label{BSRT_WS}
\end{figure}

\begin{figure}[htb]
\centering
\includegraphics[width=1.0\linewidth]{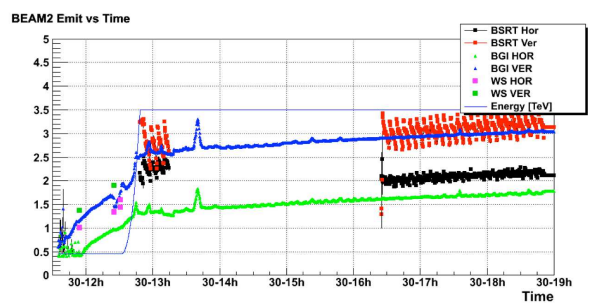} 
\caption{Evolution of the B2 beam emittance during a fill as measured by the BGI, BSRT and the wire scanner. The vertical BGI follows the BSRT and the WS while the horizontal one is quite off.}
\label{BGI}
\end{figure}

\subsection{Rest gas ionization monitor}
The BGI allows the measurement of the transverse projection of the beam in one direction (horizontal or vertical).
The particles in the beam leave behind an ionized column where the ions (free-electrons) density reproduces the density of the beam. An electric field drifts the electrons toward a multi channel plate while a magnetic field, parallel to the electric one, guides the electrons and avoids smearing due to the thermal velocity and beam space charge effects. The MCP multiplies the impinging electrons which are imaged on a phosphor screen from where it can be acquired using an intensified camera \cite{bgi}. This device is very sensitive to many effects, beam space charge, electron cloud, stray fields and fields non homogeneity, but if all parameters are well controlled the accuracy can be elevated. The problem with the BGI is that in order to obtain sufficient signal either a large number of particles in the beam is needed or a local pressure bump must be created. A local pressure bump will also increase beam losses locally imposing a stringent control and limits. Due to these constraints the four BGIs installed in LHC (1 per beam and per plane) could not be fully commissioned.  The results obtained so far show that for some device the agreement between BGI, BSRT and WS is good while for the others it is quite off. The reasons for this discrepancies will be investigated in 2011.

\section{Beam halo and tails}
In the BSRT design it is foreseen to install an optical mask in order to cut the core of the beam and allow the observation of the tails, a technique known in astronomy as a "corona" monitor used in sun observations.
At the moment the required hardware is not installed as this functionality is not considered a high priority, but if really needed this could be developed in a reasonable amount of time. The overall performance of this halo monitor is in the end limited by the amount os scattered light in the optical components and in general inside the telescope hatch.

\section{Beam position at the IP}
In order to monitor the beam position at the IP dedicated beam position monitors are installed just outside of the experiments and before the triplets. Around all the four interaction regions strip line pick-ups are installed, the choice for this type of devices is dictated by the fact that in multi bunch mode an incoming and an outgoing bunch can pass trough the detector with very small time difference making impossible to disentangle the signals from one or the other beam. The strip line devices have the advantage that although each strip has an upstream and a downstream port, the beam induces a pulse only in the upstream port, so the signals of the two beams  can be read out independently from the two ports. The disadvantage of this method is that the electronic chains used to acquire the signals are different from one beam and the other, adding the possibility of an unknown electronic offset and making the overlap of the two beams more difficult. For this reason around IP1 and IP5 additional button pick-ups have been installed, these devices have the advantage that the readout chain is the same for the incoming and the outgoing beam so that any electronics offset cancels out. The disadvantage is that the bunch spacing must be larger than 150\,ns.
In order to calculate the overlap one can use a simple ballistic model, the experiments have however strong magnetic fields which can complicate the situation, especially for LHC-b and ALICE where spectrometer magnets exist.
The orbit mode resolution for the strip line detectors is of the order of 1\,$\mu$m and for the buttons it is slightly worse, but the electronic offset can be substantially larger than this value.

\section{Conclusions}
In order to compute the luminosity of the LHC beams several parameters must be measured accurately. In particular the distribution of charges around the machine needs to be precisely known in order to calculate the fraction of colliding charges. The wall current monitor and the longitudinal density monitors are both able to provide this information, with the LDM probably able to give better accuracy, also because it can measure the DC component as well while the WCM is limited to AC. The other important parameter to measure and monitor is the transverse emittance and for this purpose the wire scanners and the BSRT are providing the required information.

\end{document}